\documentclass[a4paper,twocolumn]{Gaia2004} % European paper size
\usepackage{times}      % for font
\usepackage{epsfig}     % for figure inclusion
\usepackage{natbib}     % for bibliography
\title{Beyond the Galaxy with Gaia: evolutionary histories of galaxies in the Local group}

\author[1,2]{A.~Ku\v{c}inskas}
\affil[1]{National Astronomical Observatory of Japan, 2-21-1 Mitaka, Tokyo, 181-8588, Japan}
\affil[2]{Institute of Theoretical Physics and Astronomy, Go\v{s}tauto 12, Vilnius 01108, Lithuania}

\author[3]{L.~Lindegren}
\affil[3]{Lund Observatory, Lund University, Box 43, SE-221 00, Lund, Sweden}

\author[4]{V.~Vansevi\v{c}ius}
\affil[4]{Institute of Physics, Savanoriu 231, Vilnius 02300, Lithuania}

\bibpunct{(}{)}{;}{a}{}{,}  % to set bibliography punctuation to A&A style

\begin{document}

\keywords{Insert key words here}

\maketitle

\begin{abstract}

Gaia will play an important role in providing information about
star formation histories, merging events, intergalactic streams
etc., for nearby galaxies of the Local Group. One of the most
crucial contributions will be proper motions, especially for stars
in the outermost parts of the galaxies, obtainable for stellar
populations to $\sim150$\,kpc with RGB stars. Together with radial
velocities for the brightest giants ($<80$\,kpc), this will
provide membership information for individual stars and global
kinematical picture of the most nearby galaxies, including the
Magellanic Clouds (MCs). Gaia will also provide photometric
metallicities ($\sigma([M/H])<0.3$) for individual giants and/or
supergiants in dwarf galaxies to $\sim200$\,kpc. MSTO ages will be
possible for the youngest stellar populations in the most nearby
galaxies (e.g., MCs), whereas stars on RGB/AGB may provide age
estimates for populations to $\sim150$\,kpc. Gaia will allow to
study the outermost parts of the galaxies, which (because of their
large spatial extent) are difficult to assess from the ground.
Apart from allowing to clarify the structure and evolution of the
dwarf galaxies, this will also make it possible to investigate
galactic tidal debris, thus providing additional details for the
global picture of formation and evolution of the Milky Way Galaxy.

\end{abstract}

\section{Introduction}

The past few decades have been extremely fruitful in terms of
advancing our understanding about the Local Group (LG), both on
the level of individual galaxies, and the LG at large (van den
Bergh 2000). This was facilitated by the deployment of HST, large
ground-based telescopes (VLT, Keck, Subaru, Gemini) and new
instrumentation (especially wide field cameras and multi-object
spectrographs), crucial for studies of faint stellar populations
in the LG galaxies.

Gaia will be a powerful tool to study formation and evolution of
the Milky Way, but apart from this it will also provide a wealth
of information on individual stars (giants and supergiants) in
distant stellar populations, such as galaxies of the LG. However,
given the fast development of new observational facilities,
especially those foreseen to come into operation already within
$\sim$10 years, it is crucial to understand whether and what Gaia
may contribute in this rapidly evolving field of research in
10--15 years from now. We attempt to address this question briefly
in the forthcoming pages.

\begin{figure}[]
\begin{center}
    \leavevmode
    \centerline{\epsfig{file=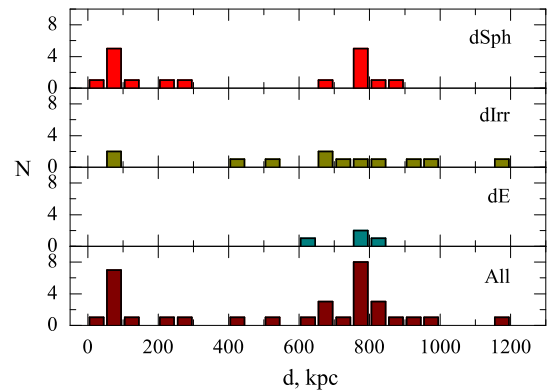,width=0.9\linewidth}}
\end{center}
\caption{Distances to Local Group galaxies (data from Grebel et
al. 2000).} \label{}
\end{figure}

\begin{table}[]
\caption{Local group galaxies within 300\,kpc from the Sun}
\label{}

\begin{center}
\leavevmode

\begin{tabular}[h]{lllcc}
\hline
\noalign{\smallskip}
Name    &    Type    & $D_{\odot}$ & $r_{\rm core}$ & $r_{\rm tidal}$ \\
        &            &     kpc     &     arcmin     &      arcmin     \\
\hline
\noalign{\smallskip}
 Sgr    &  dSph,N?   &  $28\pm3$   &       --       &     $>600$      \\
 LMC    &  IrIII-IV  &  $50\pm5$   &     \multicolumn{2}{c}{$646$}    \\
 SMC    &  IrIV/IV-V &  $63\pm10$  &     \multicolumn{2}{c}{$372$}    \\
 UMi    &  dSph      &  $69\pm4$   &     $16$     &     $51$      \\
 Dra    &  dSph      &  $79\pm4$   &     $9$      &     $28$      \\
 Sex    &  dSph      &  $86\pm6$   &     $17$     &     $160$       \\
 Scl    &  dSph      &  $88\pm4$   &     $6$      &     $77$      \\
 Car    &  dSph      &  $94\pm5$   &     $9$      &     $29$      \\
 For    &  dSph      &  $138\pm8$  &     $14$     &     $71$        \\
 Leo II &  dSph      &  $205\pm15$ &     $3$      &     $9$       \\
 Leo I  &  dSph      &  $270\pm10$ &     $3$      &     $13$      \\
\hline
\end{tabular}

\end{center}

Notes: morphological types and distances are from Grebel et al.
(2000), core and tidal radii are from Mateo (1998). Apparent
diameters at 25 magnitude isophote are given for LMC and SMC (LEDA
database, {\tt http://leda.univ-lyon1.fr/search.html}).

\end{table}

\section{The Local Group: current status and future ground-based prospects}

\subsection{Current status}

A list of nearby LG galaxies which may be accessible with Gaia is
given in Table 1 {\footnote{Note that dwarf galaxies in the LG are
clustered around the two major spirals, i.e., the Milky Way and
M31 (Fig.1). Beyond 300\,kpc and closer than 600\,kpc, there are
only two dwarf irregular galaxies (Phoenix, 400\,kpc, and
NGC~6822, 500\,kpc). NGC~6822 is a star-forming galaxy and thus
may be an interesting (and challenging) target for Gaia with
supergiant stars (see Sect.3).}}. It is obvious that a
comprehensive study of the nearby LG galaxies requires
instrumentation with high sensitivity and large field of view
(FOV), as most of them occupy large areas on the sky (note that
stellar populations of dwarf galaxies may extend to several
apparent radii, see e.g. Vansevi\v{c}ius et al. 2004, thus
covering an area of several square degrees on the sky). The demand
for high sensitivity will imply the use of largest telescopes
available; however, their FOVs are prohibitively small compared to
the spatial extent of LG galaxies, especially the nearby ones.
Thus current studies of LG dwarfs are mostly limited to small
areas in individual galaxies (typically - their central parts).

The outer edges of dwarf galaxies, however, are extremely
interesting, as they may contain important imprints about the
formation and evolution of individual galaxies, as well as their
interaction with other galaxies in the LG (spatial gradients in
star formation and chemical enrichment histories, kinematical
properties, etc.). However, field contamination becomes
increasingly important at larger radii, thus membership
information (typically based on kinematical properties) is crucial
to distinguish galactic populations from the field stars.

Current kinematical studies of LG galaxies are mostly based on
radial velocities, which are possible only for brightest giants
and supergiants with 8-m class telescopes. Several studies have
aimed recently at obtaining proper motions of individual stars in
nearby LG galaxies, but: (1) the accuracy is low, mainly due to
the relatively short time span between the measurement epochs, (2)
studies of this kind require 8-m class telescopes (or HST) to
ensure the required spatial resolution - which, given today's
situation, means small FOVs again. Thus, astrometry of LG galaxies
presently remains limited to small selected areas in the most
nearby galaxies.

Star formation histories of the nearby LG galaxies are relatively
well understood (e.g., Aparicio 2004; Grebel 2000), as their main
sequence turn-off point (MSTO) stars are well within reach with
currently available telescopes (see Fig.2, which illustrates the
possibilities with currently available instrumentation). Knowledge
is accumulating gradually about their chemical enrichment
histories too, primarily with the aid of new multi-object
spectrographs on 8-m class telescopes (e.g., Venn et al. 2004;
Shetrone et al. 2003; Tolstoy et al. 2003). Indeed, the majority
of these investigations are confined to the central parts of
individual galaxies (or several selected areas at different radial
distances), which is imposed again by the limited FOVs of
currently available instrumentation.

\subsection{Relevant future ground-based projects}

\begin{figure}[]
\begin{center}
    \leavevmode
    \centerline{\epsfig{file=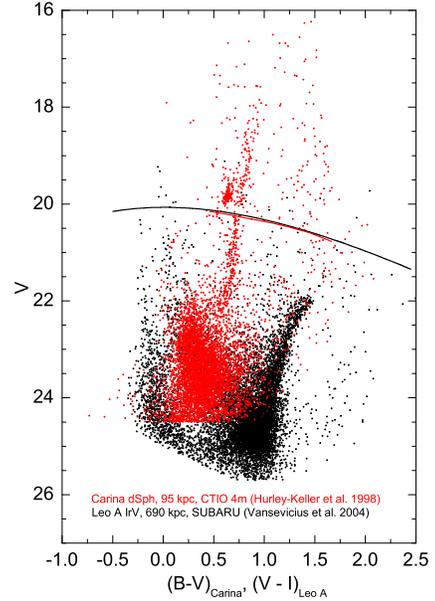,width=0.7\linewidth}}
\end{center}
\caption{Observed CMDs of Carina and Leo A dwarf spheroidals.
Solid lines indicate Gaia detection limit at $G=20$, in $V-(V-I)$ plane
(black) and $V-(B-V)$ plane (grey). Note that photometric color
indices $(B-V)$ and $(V-I)_{\rm Cousins}$ are very similar in a
given $T_{\rm eff}$ range, thus detection limits nearly coincide
in both color planes.} \label{}
\end{figure}

Considerable advances can be expected in all of these fields with
the large area surveys which will be carried out in the near
future using new wide-field telescopes (Table~2). Although far
from being complete, the list in Table~2 gives a clear indication
that the FOVs of available instrumentation will increase
dramatically in $\sim$5--10 years from now, thus a few pointings
will be sufficient to cover areas extending to several square
degrees on the sky. Combined with the high sensitivity (8-m class
telescope in case of LSST!), this will provide excellent
possibilities for studying even the outermost parts of the LG
galaxies in a routine way.

\begin{table*}[tb]
\caption{Selected future spectroscopic and photometric wide field
surveys, relevant for the studies of Local Group galaxies.}
\label{}
\begin{center}
\leavevmode
\begin{tabular}[h]{llllcclcc}
\hline \noalign{\smallskip}
Survey      &      Type      &     Limiting      &    Telescope   &         FOV        &     Photometry    &      Year     &  Webpage \\
            &                &        mag        &                &                    &                   &               &          \\
\hline \noalign{\smallskip}
\multicolumn{8}{c}{Spectroscopy}                                                                                                      \\
\noalign{\smallskip}
 RAVE       &    All-sky     &       $V=16$      &   1.2m + ?$^1$ & $6\times6$~deg     &         --        & 2006-2010$^2$ &    1     \\
\noalign{\smallskip}
\multicolumn{8}{c}{Photometry}                                                                                                        \\
\noalign{\smallskip}
 CFHTLS     &  Selected$^3$  & $i^\prime=24.4^4$ &   3.6m (CFHT)  &  $1\times1$~deg    &  Optical$^5$      & 2003-         &    2     \\
 VST        &  Selected$^a$  & $i^\prime=24.2$   &   2.6m         &  $1\times1$~deg    &  Optical$^6$      & 2005-         &    3     \\
 UKIDSS     &    Selected    &   $K=18.4^7$      &   3.8m (UKIRT) & $0.9\times0.9$~deg & NIR$^8$           & 2005-2012     &    4     \\
 VISTA      &  Selected$^a$  &   $K=20.1^9$      &   4.0m         & $1\times1$~deg     &  NIR$^{10}$       & 2006-         &    5     \\
 Pan-STARRS &    All-sky     &   $i=25.4^{11}$   & $4\times1.8$m  & $2.7\times2.7$~deg &  Optical$^{12}$   & 2006-$^{13}$  &    6     \\
 LSST       &    All-sky     &   $V=24.0^{14}$   & 8.4m           & $3.5\times3.5$~deg &  Optical$^{15}$   & 2012-         &    7     \\
\hline
\end{tabular}
\end{center}

Notes:

$^{a}$ All-sky survey possible.
$^1$ AAO Schmidt; northern hemisphere counterpart sought.
$^2$ Phase I: 2004-2005 ($10^5$ stars, $V<13$); Phase II: 2006-2010 ($5\cdot10^7$ stars, complete to $V=16$).
$^3$ MegaPrime/MegaCam on CFHT is available for PI programs.
$^4$ Very Wide survey; deeper for Deep Synoptic and Wide Synoptic surveys.
$^5$ SDSS u$^\star$ g$^\prime$ r$^\prime$i$^\prime$ z$^\prime$.
$^6$ SDSS u$^\prime$ g$^\prime$ r$^\prime$i$^\prime$ z$^\prime$, Johnson $BV$, Str\o mgren $v$, H$\alpha$.
$^7$ Large Area Survey (NIR counterpart to SDSS); deeper for smaller area surveys.
$^8$ Broad-band: Johnson $JHK$ + $Y$; narrow-band: H$_2$1-0~S1, Br-g (Tokunaga et al. 2002).
$^9$ 5$\sigma$, 15-min exposure limit.
$^{10}$ $zJHK$; optical camera foreseen in the original proposal too, subject to funding.
$^{11}$ 60 min, 5$\sigma$.
$^{12}$ $griz$ + $y,w$.
$^{13}$ First light from the first prototype telescope scheduled for 2006.
$^{14}$ Single exposure (10 sec), 10$\sigma$.
$^{15}$ $griz$ + $Y$.

Webpages:

(1) ${\tt http://www.aip.de/RAVE/}$ ; ${\tt
http://www.aao.gov.au/ukst/6dF\_RAVE.html}$

(2) ${\tt http://www.cfht.hawaii.edu/Science/CFHTLS/}$

(3) ${\tt http://twg.na.astro.it/vst/vst\_homepage\_twg.html}$

(4) ${\tt http://www.ukidss.org/}$

(5) ${\tt http://www.vista.ac.uk/}$

(6) ${\tt http://pan-starss.ifa.hawaii.edu/project/}$

(7) ${\tt http://www.lsst.org/lsst\_home.shtml}$

\end{table*}

\section{The Local Group with Gaia: morphological and evolutionary histories
of the nearby galaxies?}

Obviously, an important contribution of Gaia will be proper
motions (precise parallaxes of late type giants, $\sigma(\pi)/\pi
\le 0.1$, will not be available beyond $\sim10$\,kpc). Proper
motions of individual giants and supergiants in LG galaxies will
allow to discriminate between the member and field stars
(Table~3). However, the accuracy may not be sufficient for
detailed kinematical studies of individual galaxies (which will
require the precision in proper motions of $1-2$\,km/s - see,
e.g., Wilkinson et al. 2002). Note however, that proper motions of
individual stars in LG galaxies will be also feasible with future
wide-field instrumentation (e.g., with PanSTARSS), with the time
spans between the measurement epochs of $\sim5-10$\,years possible
already in 2010--2015 (though the accuracies will be lower than
those obtained with Gaia).

\begin{figure}[!hb]
\begin{center}
    \leavevmode
    \centerline{\epsfig{file=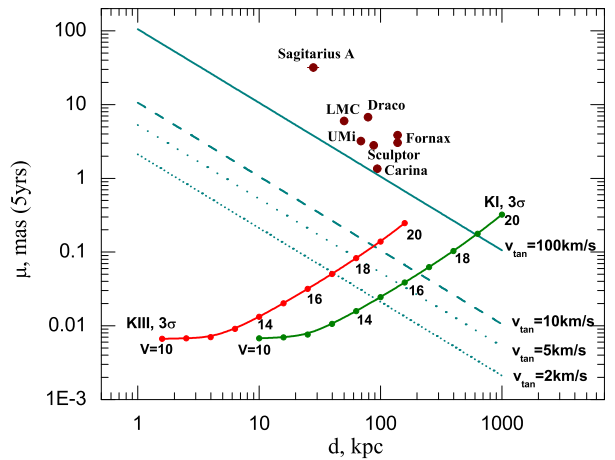,width=1.0\linewidth}}
\end{center}
\caption{Expected error in proper motion with Gaia (solid lines)
for KIII star ($M_V=-1.0$, RGB/early-AGB tracer) and KI star
($M_V=-5.0$, red supergiant). Errors are given at $3\sigma$ level;
numbers are apparent $V$ magnitudes. Diagonal lines indicate
proper motions corresponding to different tangential velocities.
Dots indicate LG galaxies with measured proper motions. All proper
motions are given for a 5-year lifetime of the Gaia mission.}
\label{}
\end{figure}

Radial velocities will be considerably more difficult with Gaia,
as sufficient precision may be achieved only for stars brighter
than $V\sim17$, or up to the distances of $\sim40$\,kpc (Table~3;
$\sim$80\,kpc with AGB stars, $M_V=-2.5$). Indeed, bright
supergiants may be accessible at considerably larger distances
(Table~3), however, they will be available only in a few galaxies,
as stellar populations in the majority of nearby LG galaxies are
old, showing no signs for recent star formation (Aparicio 2002).
Moreover, their numbers will be too scarce to perform a detailed
kinematical analysis.

\begin{table}[]
\caption{Predicted accuracies (versus $V$ magnitude) for parallax
($\pi$), proper motion ($\mu$), and radial velocity ($v_{\rm r}$)
with KIII and KI tracers (all for 5 year mission lifetime).
Relative parallax error is not shown when $\sigma(\pi)/\pi>1$.}
\label{}

\begin{center}
\leavevmode

\begin{tabular}[]{llccccccc}
\hline \noalign{\smallskip}
  $V$  &  $d$  & $\sigma(\pi)/\pi$ & $\sigma(\mu)$ & $\sigma(v_{\rm r})$ \\
       &  kpc  &                   &       km/s    &        km/s         \\
\hline \noalign{\smallskip}
\multicolumn{5}{c}{KIII, $M_V=-1.0$}                                     \\
\noalign{\smallskip}
 15    &  16   &        0.20       &      0.1      &         1.7         \\
 16    &  25   &        0.50       &      0.3      &         4.1         \\
 17    &  40   &         --        &      0.6      &        10.3         \\
 17.5  &  50   &         --        &      1.1      &        23.4         \\
 18    &  63   &         --        &      1.7      &     $>$35           \\
 19    &  100  &         --        &      4.4      &         --          \\
 20    &  160  &         --        &      12.3     &         --          \\
\noalign{\smallskip}
\multicolumn{5}{c}{KI, $M_V=-5.0$}                                       \\
\noalign{\smallskip}
 15    &  100  &         --        &      0.8      &         1.7         \\
 16    &  160  &         --        &      1.9      &         4.1         \\
 17    &  250  &         --        &      5.0      &        10.3         \\
 17.5  &  320  &         --        &      7.8      &        23.4         \\
 18    &  400  &         --        &     13        &     $>$35           \\
 19    &  630  &         --        &     35        &         --          \\
 20    & 1000  &         --        &    100        &         --          \\

\hline
\end{tabular}

\end{center}

Note: based on the astrometric accuracies from de Bruijne (2003),
and accuracies in radial velocities from Katz et al. (2004).

\end{table}

Gaia will also provide precise astrophysical parameters for a
number of individual stars in nearby LG galaxies, from the
medium-band photometry (see Jordi \& H{\o}g, this volume). This
will be indeed an important contribution since so far neither
spectroscopy nor medium-band photometry is planned with the deep
ground-based wide-field surveys (note though, that the former may
be possible with the prime-focus multi-object spectrographs on 8-m
class telescopes in the near future, e.g., FMOS on SUBARU; and the
implementation of the latter is a straightforward task). The
availability of precise astrophysical parameters may also allow to
study star formation histories of individual galaxies using
RGB/AGB stars (Ku\v{c}inskas et al. 2003).

To a certain extent, however, at least some of these tasks may be
accomplished with future wide-field telescopes (e.g., PanSTARSS).
With their high sensitivity and large FOVs, these instruments will
be serious competitors for Gaia, especially if backed-up with the
capability for multi-color medium-band photometry and multi-object
spectroscopy. However, the superiority of Gaia will be in the high
accuracy and homogeneity of its data. Gaia will provide precise
information both on kinematical and astrophysical properties of
individual stars, obtained with a single, stable and
well-calibrated instrument. Moreover, the data will be acquired in
the same instrumental system as for the millions of stars in the
Milky Way. This will deliver an unique and homogeneous data set,
which will allow to perform a comprehensive and in-depth analysis
of the formation and evolution of the LG galaxies.

\section*{Acknowledgments}

AK acknowledges Research Fellowship of the National Astronomical
Observatory of Japan. This work was also supported by the
Wenner-Gren foundations.

\end{document}